\documentclass{PoS}
\usepackage{parskip}
\def\lh{\hat{L}}
\def\eh{\hat{E}}
\def\qh{\hat{Q}}
\def\uh{\hat{U}}
\def\dh{\hat{D}}

\title{FCNC Decays of the Top Quark}

\ShortTitle{FCNC Decays of the Top Quark}

\author{\speaker{Debjyoti Bardhan}\thanks{The speaker wishes to thank all the other authors
on the paper. He would especially like to thank Prof. Sreerup Rauychaudhuri and Dr. Diptimoy Ghosh for their help with this talk. }\\
        Tata Institute of Fundamental Research\\
        E-mail: \email{debjyoti@theory.tifr.res.in\\
            TIFR/TH/17-14}}

\abstract{If new physics (e.g. SUSY) does not show up as direct evidence at the
LHC, it could still be observable in FCNC processes involving the
$t$-quark. We take a close look at the process $t\to  c + h/Z$ and show that
its branching ratio in the Standard Model is subject to three
mechanisms of suppression. To obtain an observable signal, one needs
to evade all these mechanisms in a theory beyond the Standard
Model. We show that a theory like the cMSSM cannot provide a big
enough enhancement. However, in a framework like $R$-parity-violating
SUSY, observable signals are a distinct possibility.}

\FullConference{9th International Workshop on the CKM Unitarity Triangle\\
		28  November - 3 December 2016\\
		Tata Institute for Fundamental Research (TIFR), Mumbai, India}

\begin{document}

\section{Introduction}
In the search for New Physics (NP), much effort has been devoted to 
direct searches. Particles at the LHC can be produced either in 
pairs, or resonantly, or in associated with a Standard Model (SM)
particle. However, the energy reach of the LHC is $\sim 2.5 {\rm \ TeV}$, 
which is roughly the scale at which a significant number 
of hard interactions take place. If direct searches fail, it makes
sense to take a look at indirect searches and exploring Flavour 
Changing Neutral Current (FCNC) decays of the top quark 
is one avenue. 

The top quark is the heaviest fundamental particle discovered and,
unlike other quarks in the SM, it decays before hadronising. Since
phenomena like hadronisation are inherently 
low energy strong coupling effects and an accurate measure of them 
isn't known, the decay of the top quark, being devoid of 
these effects, is theoretically clean. 

The top quark decays into a $b$-quark and a $W$-boson most of the 
time, but very rarely also decays into an up-type quark, like a 
$c$ or a $u$-quark, associated with a neutral boson like a $Z$ 
or a Higgs boson. These rare decay modes have very small SM branching
ratios (BRs) ($\sim10^{-15}$) and are beyond the detection capabilities of
the LHC, which can optimistically probe a maximum of $10^{-5}$. We 
can study the different methods of suppression because of which 
the BR in the SM is so small; consequently, we can think about 
a scenario by which the BR can be enhanced. The enhanced BR can be 
detected at the LHC especially because the SM BR being so small 
won't swamp out the signal. 

\section{Modes of Suppression}
Flavour changing neutral current (FCNC) decays of the top 
quark are extremely suppressed in the SM and any deviation from this may be identified in 
colliders as signatures for New Physics. We performed a study of such FCNC decays in \cite{Bardhan:2016txk}.

\begin{figure}[htb!]
\centering
\includegraphics[scale=0.7]{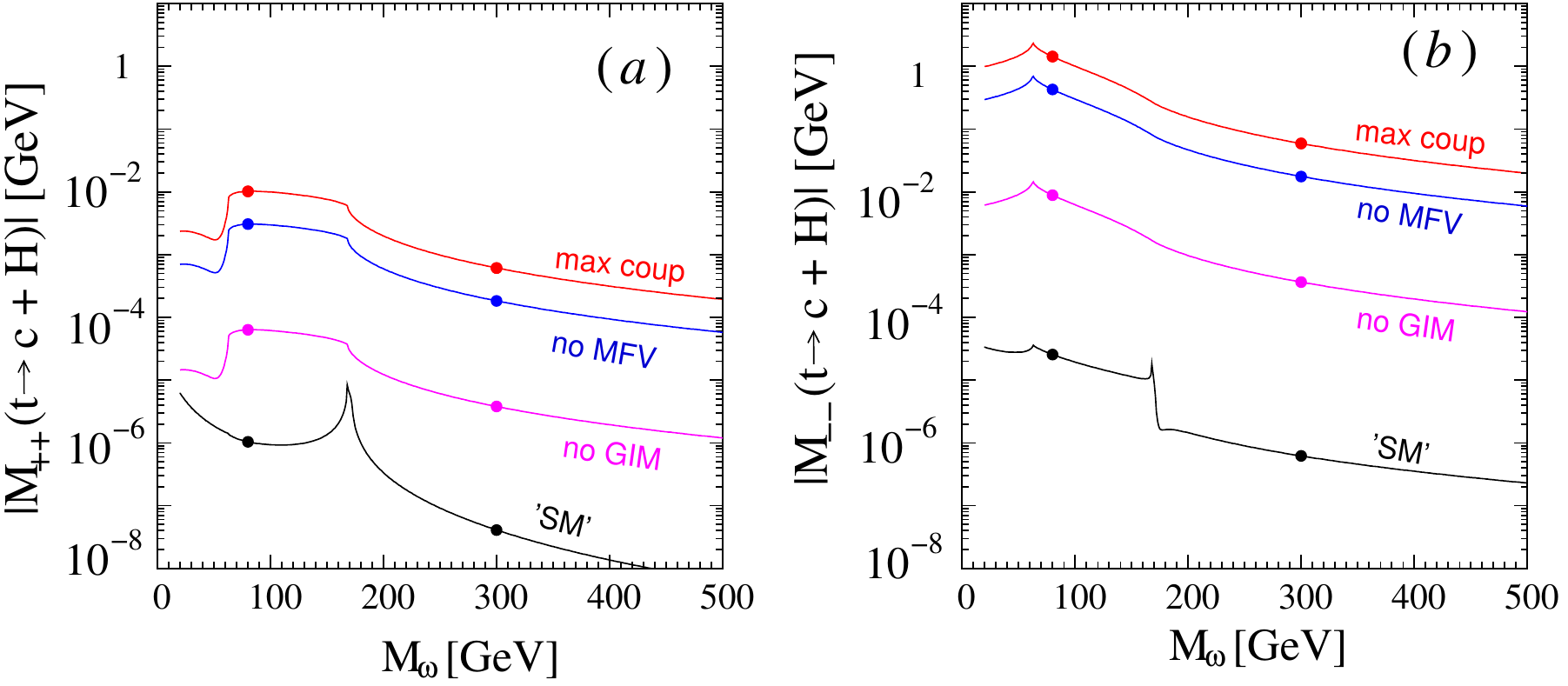}
\caption{{\small{\it The absolute value of the two helicity amplitudes and how the 
relaxation of the different suppression factors enhances the amplitude. The subscripts 
of $\mathcal{M}$, plotted on the y-axis, denotes the helicity of the charm and the top.}}}
\label{fig:toyamps}
\end{figure}

The different mechanisms of suppression that lead to the tiny branching ratios of these FCNC processes are: 
\begin{itemize}
\item The GIM suppression in these decays
\item The Minimal Flavour Violation (MFV) framework which leads to a hierarchy of the
values of the CKM matrix elements 
\item The smallness of the weak coupling constant 
\end{itemize}
In order to study the effects of these modes of suppression in detail, consider a toy
model where, apart from the quarks and the Higgs ($H$), there is a flavour changing charged scalar field ($\omega$) and the coupling strength between $\omega$ and $H$ is given by $\xi$. The 
interaction Lagrangian is given by:
\begin{equation}
\mathcal{L}_{int} = \xi \omega^+ \omega^- H + \sum_{i,j = 1}^3 \left(\eta V_{ij} \bar{u}_{iL} d_{jR} \omega^+ + h.c.\right)
\end{equation}
A similar Lagrangian can be written down for a theory with a $Z$ boson, instead of the Higgs.
This is a theoretical laboratory for studying the effects of relaxing each mode of 
suppression, one at a time. The effect on the amplitude is summarised in Fig. \ref{fig:toyamps}.
 
The effects on the amplitude can be summarised as follows: 
\begin{itemize}
\item {\bf GIM Mechanism:} The unitarity of the CKM matrix leads to this mode of suppression, first shown in \cite{Glashow:1970gm}. The matrix element for the process for a process in which one quark $q$ changes to another quark $q'$ of the same charge can be written as 
\begin{equation}
M_{qq'} = \sum_{i=1}^3 V^\ast_{qi} V_{q' i} A\left(x_i, M_W\right) = \sum_{i=1}^3 \lambda_i A\left(x_i, M_W\right)
\label{eqn:gim_amplitude}
\end{equation}
where $x_i = m_i^2/M_W^2$ carries the generation information, $M_W$ is the scale of the
charged current interaction and $\lambda_i = V^\ast_{qi} V_{q' i}$. This can be expanded as 
\begin{equation}
A(x_i,M_W) = A_0(M_W) + x_i A_i'(M_W) + \frac{1}{2} x_i^2 A_i''(M_W) + \ldots
\end{equation}
where primes represent differentiation with respect to $x_i$. 
Putting this back in \ref{eqn:gim_amplitude} and using the fact that 
$\sum_i \lambda_i = 0$, due to the unitarity of the CKM matrix, it can be seen that the
 dominant term is actually not proportional to $A_0$, but to $x_i A_i'$. This gives us 
 a suppression of $x_i$ in the amplitude; for the top to charm decay, 
 $x_i \sim x_b \sim m_b^2/M_W^2 \sim 10^{-3}$, presenting a suppression of 
 $\sim 10^{-6}$ in the decay width. 

\item {\bf MFV Framework:} The MFV framework \cite{Buras:2003jf} means that the CKM matrix has a strong hierarchy with regards to the value
of its elements. In our case, of a top decay to a charm, the dominant CKM matrix element involved
is $V_{tb}^\ast V_{cb} \sim 0.04$. An alternative prescription, which preserves unitarity but
doesn't have thehierarchy seen here, can be constructed. For example, we might have
\begin{equation}
V = \left(\begin{array}{ccc}
1 & 0 & 0 \\
0 & \cos \theta & \sin \theta \\
0 & -\sin \theta & \cos \theta \end{array} \right) 
\end{equation}
The dominant CKM element here is $\lambda_3 = \sin \theta \cos \theta = 0.5 \sin 2\theta$, which can take up a maximum value of $0.5$, as opposed to the 0.04 as in the previous case. 

\item {\bf Coupling Constant:} Finally, the weak coupling constant is rather small in magnitude and we can think 
upon a new physics model which can have large couplings. In our toy model, we can get a modest 
enhancement of $\sim 7$. 
\end{itemize}
The overall enhancement in the amplitude is then $10^3 \times 25 \times 7 \sim 1.75 \times 10^5$ and thus, in the branching ratio, it is a factor of $\sim 3 \times 10^{10}$.
\section{Results in different models}
\subsection{Standard Model}
The SM BR is calculated to be $5 \times 10^{-15}$. Violating all the three suppression mechanisms, one can hope to reach a BR of $~10^{-5}$. 

\subsection{cMSSM} 
In the constrained Minimal SuperSymmetric Model (cMSSM), given by the four parameters
 - the universal fermion mass scale at high energy $m_{1/2}$, the universal scalar mass scale at
  high energy $m_0$, the trilinear Higgs coupling parameter $A_0$ and the sign of the higgsino
  mass parameter $sgn(\mu)$ - there exist charged Higgs bosons which naturally violate the GIM
   mechanism. 

\begin{figure}[t]
\centering
\includegraphics[scale=0.4]{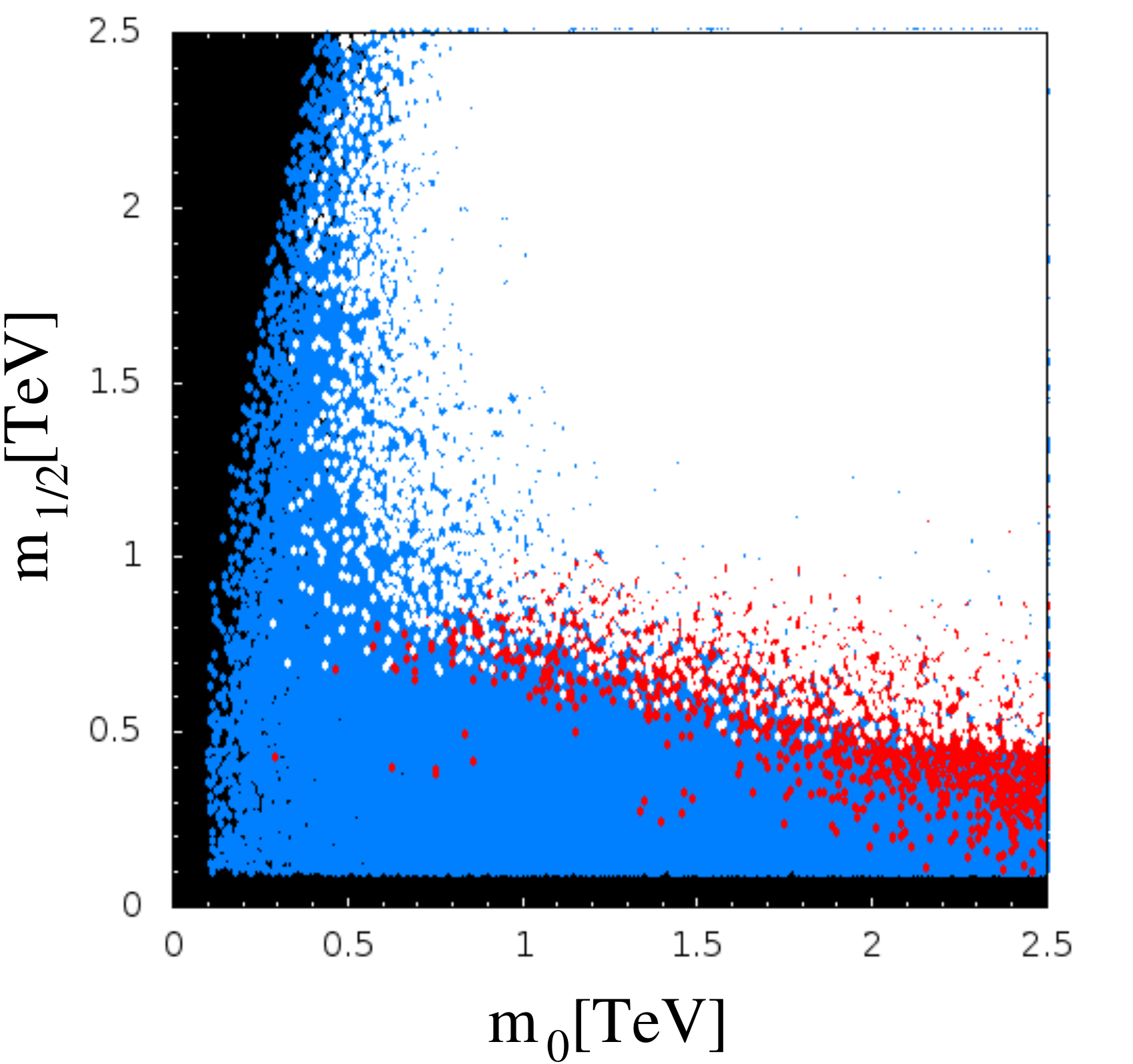}
\includegraphics[scale=0.4]{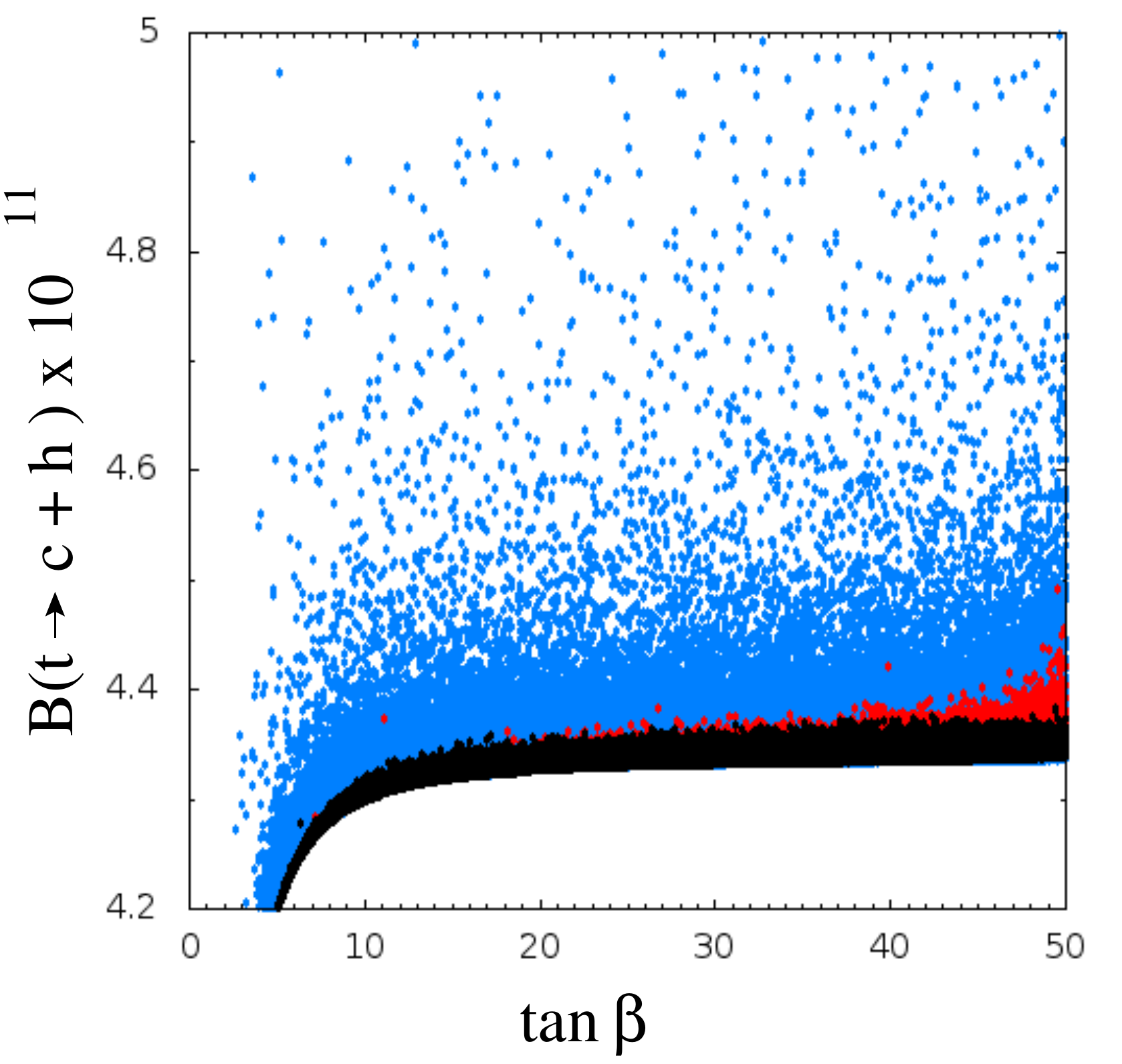}
\caption{{\small{\it The cMSSM parameter space spanned by $m_{1/2}$-$m_0$. The black points
are excluded from theoretical considerations, like unstable vacuum, tachyonic states etc. 
The blue points are excluded by the Higgs mass constraint, while the red points are 
excluded from flavour constraints. The plot on the right is the branching ratio for these 
points, where the colours are the same, except that the white (allowed) points on the left   correspond to the black points on the right.}}}
\label{fig:cmssm_cons}
\end{figure}
However, the four parameters cannot take any value - they are constrained by various experimental
inputs, most notably the Higgs mass. The effect of various constraints can be estimated from Fig \ref{fig:cmssm_cons}. 

These constraints are responsible in raising the mass of the 
charged Higgs bosons in the theory, which suppresses the amplitude significantly. Any 
enhancement achieved by breaking GIM is offset partially by the heavy charged Higgs in the theory.
The cMSSM, furthermore, doesn't go beyond the MFV framework and the enhancement due to the coupling is also not very large. Overall, the cMSSM enhancement doesn't exceed $\sim 10^4$ in the branching ratio. 

  \begin{figure}[t]
\centering
\includegraphics[scale=0.7]{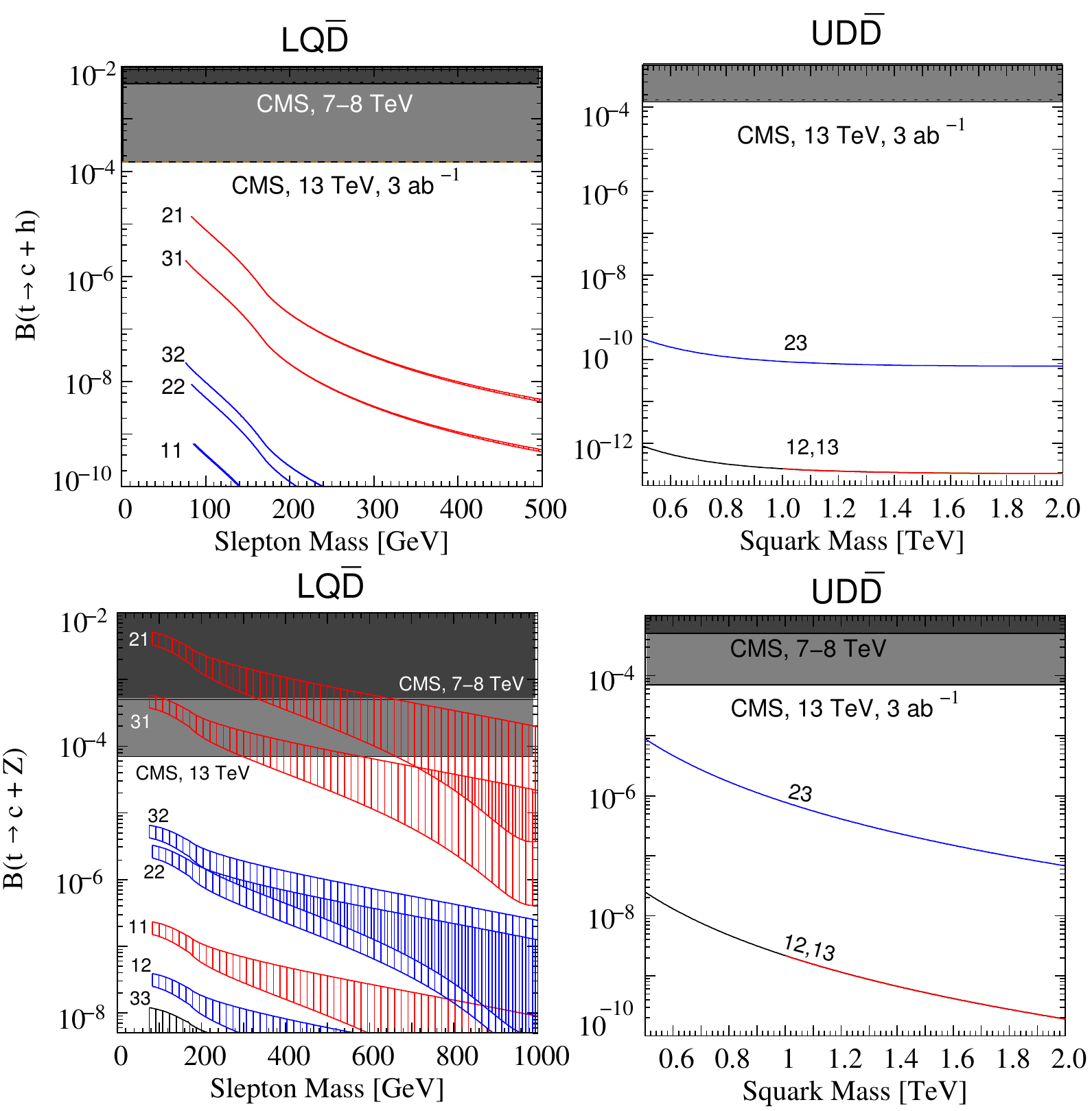}
\caption{{\small{\it Illustrating the variation in the branching ratios $B(t \to c + h_0 )$
(upper panels) and $B(t \to c + Z_0 )$ (lower panels) with increase in the sfermion masses.
For the panels on the left, which show branching ratios proportional to $(\lambda_{i2k}
\lambda_{i3k})^2$ with the values of $ik$ marked next to each curve, the mass of the
slepton $e_{Li}$ is plotted along the abscissa, and the mass of the squark $\tilde{d}_{Rk}$
is responsible for the thickness of the lines in the upper panel and the hatched region in
the lower panel. For the panels on the right, which show branching ratios
proportional to $(\lambda_{2jk} \lambda_{3jk})^2$ with the values of $jk$ marked next to 
each curve, the mass of the squark $\tilde{d}_{Rk}$ is plotted along the abscissa. The dark
(light) grey shaded regions represent the experimental bounds (discovery limits) from the 
LHC, operating at $7 - 8$ TeV (13 TeV, projected}}}
\label{fig:tcZ_rpv}
\end{figure}
\subsection{RPV SUSY}
 SUSY models which break $R$-parity (a  good review of RPV-SUSY is \cite{Barbier:2004ez}) are particularly interesting in this context
for several reasons: firstly, there is no unitary CKM-like mixing matrix, thus there is no GIM
 suppression; secondly, there is no MFV framework to subscribe to either, and thus there is no
  hierarchy; finally, several of the $R$-parity violating couplings can be rather large, despite
   what is commonly understood. 

The RPV superpotential is given as:
\begin{equation}
W_{\not{R}_p} = \sum_{i,j,k=1}^3 \left(\frac{1}{2} \lambda_{ijk} \lh_i \lh_j \eh^c_k + \lambda_{ijk}' \lh_i \qh_j \dh^c_k + \lambda_{ijk}'' \uh_i^c \dh_j^c \dh^c_k \right)
\label{eqn:WRPV}
\end{equation}
where each of the fields with a hat ($\hat{}$) on top represents a superfield. $\lh$ and 
$\qh$ are SU(2) superfields, while $\eh$, $\uh$ and $\dh$ are singlets. We are interested 
in the $LQD$ term (second term) and the $UDD$ term (third term) in the Lagrangian. The second 
term violates lepton number (L) and the third term violates baryon number (B). We can only 
consider either set of coupling at once, but not both together, as that leads to proton 
decay. 

The results are rather encouraging for the $t \to c Z$ decays, where we do hit the 
projected experimental bounds, as shown in Fig.~\ref{fig:tcZ_rpv}
%

\end{document}